\documentclass[twocolumn,eqsecnum,showkeys,showpacs,aps]{revtex4}

\usepackage{amsfonts,amssymb,epsfig}
\usepackage{graphicx}
\usepackage{dcolumn}
\usepackage{amsmath}

\newfont{\cyr}{wncyr10 at 9pt}
\newfont{\cyrit}{wncyi10 at 9pt}

\begin{document}

\title{Twistor representation of null two-surfaces}%

\author{Kostyantin Ilyenko}%
 \thanks{Also at the Faculty of Physics, Kharkiv National University.}%
 \email{kost@ire.kharkov.ua}
\affiliation{Institute for Radiophysics and Electronics, NAS of
             Ukraine, 12~Acad.~Proskura Street, Kharkiv 61085, Ukraine}%

\date{May 4, 2001}%

\begin{abstract}
We present a twistor description for null two-surfaces (null
strings) in 4D~Minkowski space-time. The Lagrangian density for a
variational principle is taken as a surface-forming null bivector.
The proposed formulation is reparametrization  invariant and free
of any algebraic and differential constraints. The spinor
formalism of Cartan--Penrose allows us to derive a non-linear
evolution equation for the world-sheet coordinate
$x{}^{a}(\tau,\sigma)$. An example of null two-surface given by
the two-dimensional self-intersection (caustic) of a null
hypersurface is studied.
\end{abstract}

\maketitle


\section{Introduction}
\label{Intro} The study of massless (null) objects in 4D~Minkowski
and curved space-times has drawn considerable attention over the
recent years
\cite{Brink1,Hughston-Shaw,Bandos-Zheltukhin2,Kar,Gusev-Zheltukhin,Lousto-Sanchez,Dabrowski-Larsen,Frittelli_et_al.}.
These investigations are concerned not only with physical
implications stemming from the theory of massless particles and
string theory but also with the geometrical entities which found a
convenient representation as extended null objects. The research
is mainly confined to one-dimensional null objects (massless
particles and super-particles)
\cite{Brink2,Brink1,Shirafuji,Volkov-Zheltukhin,Soroka-et.al.} and
to the null hypersurfaces because of their relevance in relativity
\cite{SST,Kossowski,Frittelli_et_al.,Ashtekar_et_al.,Dreyer_et_al.}.
It is surprising that a study of generic null two-dimensional
surfaces is quite rare (see, however,
Refs.~\cite{Penrose_97,Duggal-Bejancu}). This situation is rather
unfortunate because a generic null two-surface corresponds to the
notion of a tensionless string, which plays an important role in
the current research on the string
theory~\cite{Gusev-Zheltukhin,Zheltukhin2,Lousto-Sanchez,Zheltukhin-Linstroem}.
Besides that, null two-surfaces can naturally arise as
two-dimensional caustics of null hypersurfaces and the
availability of such a description could provide additional
insights into the geometry of the latter ones. Finally, our
understanding of the geometry of the null submanifolds in
space-times of special and general relativity is certainly
incomplete without a satisfactory description of the null
two-surfaces.

Initially, the notion of a null two-surface was put forward by
Schild~\cite{Schild} in the form of a geodesic null string, i.e. a
two-dimensional degenerate submanifold of 4D~Minkowski or curved
space-time ruled by null geodesics. The degenerate property of the
induced metric can be written in the form
\begin{equation}
\dot{x}^2\acute{x}^2 - (\dot{x}\acute{x})^2 = 0.
\label{NullDet}
\end{equation}
Here $x^a (\tau,\sigma)$ is the world-sheet coordinate,
$\dot{x}\acute{x}$ stands for $\dot{x}{}^a\acute{x}_a$, etc. The
dots and primes denote differentiation with respect to $\tau$ and
$\sigma$, respectively. It is worth noting that the degenerate
property (\ref{NullDet}) is manifestly reparametrization invariant
while the Schild's variational principle does not possess this
feature.

In the paper~\cite{Bandos-Zheltukhin2} Bandos and Zheltukhin
proposed a spinor version of the null string action functional. A
study of null string dynamics in external fields in 3D and
4D~Minkowski space-times was undertaken in
Refs.~\cite{Zheltukhin,Ilienko1,Ilienko2-Zheltukhin,Disser}. In
Refs.~\cite{Ilienko1,Ilienko2-Zheltukhin} Ilyenko and Zheltukhin
showed that interaction with antisymmetric tensor gauge field can
lead to the violation of the geodesic property of the resulting
null two-dimensional submanifold of the 4D~Minkowski space-time.

A different approach was employed in the
articles~\cite{Stachel,Gusev-Zheltukhin}. The idea was to use an
algebraically special differential two-form obeying certain
integrability conditions. In Ref.~\cite{Stachel} Stachel took a
null bivector field $p_{ab}(x)$ ($p_{ab}=-p_{ba}$,
$p_{ab}{}^*p^{ab} = 0$, $p_{ab}p^{ab} = 0$ and
$p_{a[b}\nabla_cp_{de]}$ = $0$; the star denotes dualization) as
the Lagrangian density and showed that Schild's null string could
be treated in this way. Recently, Gusev and
Zheltukhin~\cite{Gusev-Zheltukhin} have used a fundamental result
of spinor calculus on the representation of a real null bivector
in 4D~Minkowski space-time to cast the variational principle in
the form
\begin{equation*}
S = \int (\bar{\pi}_{A}\bar{\pi}_{B}\epsilon_{A'B'} +
        \pi_{A'}\pi_{B'}\epsilon_{AB}) dx^{AA'} \wedge dx^{BB'}.%
\end{equation*}
They proved the degenerate property of the resulting
two-dimensional manifold and treated the case of geodesic null
string.

In the present article we show that the above variational
principle admits a natural twistor form. The corresponding
Euler-Lagrange equations possess solutions not only in the form of
ruled null two-surfaces (geodesic null strings) but also generic
(i.e. non-geodesic, cf. Ref.~\cite{Ilienko2-Zheltukhin}) null
strings. A non-linear counterpart of the geodesic evolution
equation for a generic null string is derived.

An outline of the paper is as follows. In Sec.~\ref{VarPr} we
propose a twistor variational principle for null two-surfaces in
4D~Minkowski space-time. Next section is devoted to a study of the
corresponding equations of motion. An evolution equation for a
generic non-geodesic null two-surface is derived in
Sec.~\ref{Sec.III}. Sec.~\ref{Sec.IV} contains an example of
non-geodesic null two-surface as a two-dimensional caustic of a
wave-front. Discussion and outlook are presented in the final
section.

The conventions are those of Penrose--Rindler~\cite{SST}.
\section{Variational principle}
\label{VarPr}%
We begin with the Stachel's variational principle. By definition,
a bivector $p_{ab}(x)=-p_{ba}(x)$ is simple if the condition
$\det{(p_{ab})}=0$ holds. In 4D~Minkowski space-time one can show
that $\det{(p_{ab})}=(1/16)(p_{ab}{}^*p^{ab}){}^2$ (see
Ref.~\cite[Vol.~1]{SST}), where ${}^*p^{ab} =
(1/2!)\varepsilon^{abcd}p_{cd}$ ($\varepsilon_{0123} =
-\varepsilon^{0123} = 1$) is the dual of $p_{ab}$. This means that
there exists a pair of vector fields $u_{a}(x)$ and $v_{b}(x)$
which obey the identity $p_{ab} = u_{[a} v_{b]}$. The null
property $p_{ab}p^{ab}$~= $0$ gives
\begin{equation}
u_a u^a = 0, \hspace{1.5em} v_a v^a < 0
\end{equation}
and without loss of generality we shall assume that they are
normalized by the conditions
\begin{equation}
u_a v^a = 0, \hspace{1.5em} v_a v^a = -2.%
\label{U_VNorm}
\end{equation}
Let the spinors $\bar{\pi}^A$ and $\bar{\eta}^A$ constitute a
normalized
Newman--Penrose dyad (spin-frame) and spinor $\bar{\pi}^A$ be
chosen as to represent the coincident principal null directions of
the null bivector $p_{ab}$. Then, one  can write the following
representation for $u_a$ and $v_b$
\begin{equation}
u_a = \bar{\pi}_A\pi_{A'}, \hspace{1.5em} %
v_b = \bar{\pi}_B\eta_{B'} + \pi_{B'}\bar{\eta}_B.%
\label{Normals}
\end{equation}

Introducing a null twistor $Z^\alpha \stackrel{\mathrm{def}}{=}
(\bar{\omega}^A , \pi_{A'})$ and its complex conjugate
$\overline{Z}_\alpha = (\bar{\pi}_A , \omega^{A'})$, where
$\bar{\omega}{}^A$ is given by the usual definition
$\bar{\omega}^A = ix^{AA'}\pi_{A'}$, we obtain
\begin{equation}
iu_a dx^a  =  i\bar{\pi}_A \pi_{A'} dx^{AA'}%
           =  \overline{Z}_\alpha dZ^\alpha.%
\label{Technical2.1}
\end{equation}
The null property of the twistor $Z^\alpha$ corresponds to the
Hermitian property of $x^{AA'}$ and reflects the reality condition
imposed on the points of 4D~Minkowski space-time.

Next we consider another one-form
\begin{eqnarray}
iv_b dx^b & = & i(\bar{\pi}_B \eta_{B'} +
\pi_{B'}\bar{\eta}_B)dx^{BB'}
           \nonumber \\
     & = & \frac{i}{2}\left[(\bar{\pi}_B +
               \bar{\eta}_B)(\pi_{B'} + \eta_{B'})\right.
           \label{Technical2.2} \\
     & - & \left.(\bar{\pi}_B - \bar{\eta}_B)(\pi_{B'} -
               \eta_{B'})\right]dx^{BB'}%
           \nonumber
\end{eqnarray}
and introduce a second null twistor $W^\beta
\stackrel{\mathrm{def}}{=} (\bar{\xi}^B, \eta_{B'})$ and its
complex conjugate $\overline{W}_\beta = (\bar{\eta}_B, \xi^{B'})$.
Calculating $d(\bar{\xi}^B \pm \bar{\omega}^B)$ = $i(\pi_{B'} \pm
\eta_{B'})dx^{BB'}$ + $ix^{BB'}d(\pi_{B'} \pm \eta_{B'})$, we find
\begin{widetext}
\begin{eqnarray}
i(\bar{\pi}_B + \bar{\eta}_B)(\pi_{B'} + \eta_{B'})dx^{BB'} & = &
{}\hphantom{-}
  (\bar{\pi}_B + \bar{\eta}_B)d(\bar{\xi}^B + \bar{\omega}^B) +
  (\xi^{B'} + \omega^{B'})d(\pi_{B'} + \eta_{B'}), \nonumber \\
i(\bar{\pi}_B - \bar{\eta}_B)(\pi_{B'} - \eta_{B'})dx^{BB'} & = &
  -(\bar{\pi}_B - \bar{\eta}_B)d(\bar{\xi}^B - \bar{\omega}^B) -
  (\xi^{B'} - \omega^{B'})d(\pi_{B'} - \eta_{B'}). 
\end{eqnarray}
\end{widetext}
Subtracting these two equations and using (\ref{Technical2.2}) we
\linebreak obtain
\begin{equation}
iv_b dx^b 
          = \overline{Z}_\beta dW^\beta + \overline{W}_\beta dZ^\beta.
\end{equation}

Since
\begin{equation} p_{ab}dx^a \wedge dx^b  =  u_{[a}v_{b]}dx^a \wedge dx^b =
u_a dx^a \wedge v_b dx^b,%
\label{P_ab}
\end{equation}
we can take the following expression:
\begin{equation}
S = \int \overline{Z}_\alpha dZ^\alpha \wedge
\left( \overline{Z}_\beta dW^\beta + \overline{W}_\beta dZ^\beta
\right) \label{Action}
\end{equation}
as a twistor variational principle for the null two-surfaces. The
two-form in (\ref{Action}) is understood to be restricted to a
two-dimensional submanifold of 4D~Minkowski space-time
parametrized by $\tau$ and $\sigma$. The null property of the
twistors $Z^\alpha$ and $W^\alpha$ leads to the following
identities:
\begin{equation}
\overline{Z}_\alpha Z^\alpha = \overline{W}_\alpha W^\alpha =
\overline{Z}_\alpha W^\alpha = \overline{W}_\alpha Z^\alpha = 0.
\end{equation}
The Lagrangian density of the twistor action functional
(\ref{Action}) is multiplied by the factor $q^2$ under the gauge
transformations of the form
\begin{equation}
Z^\alpha \rightarrow qZ^\alpha , \hspace{1.5em} W^\alpha
\rightarrow q^{-1}W^\alpha + pZ^\alpha .
\end{equation}
Here $q(\tau ,\sigma)$ is a nowhere vanishing real-valued function
and $p(\tau,\sigma)$ is an arbitrary com\-plex-valued function.
This is admissible freedom for a differential form representing a
surface~\cite{Schouten}. It gives rise to the invariance of the
Euler-Lagrange equations under the above mentioned
transformations. The invariance corresponds to the possibility of
rescaling with real multiples of the extent of the null direction
tangent to the null two-surface and to addition of any real
multiple of the null direction to the space-like tangent direction
\begin{equation}
\bar{\pi}^A \rightarrow q\bar{\pi}^A, \hspace{1.5em} \bar{\eta}^A
\rightarrow q^{-1}\bar{\eta}^A + p\bar{\pi}^A.%
\label{Scale_Add}
\end{equation}
These transformations comprise the null-rotations and
boost-rotations (cf. Ref.~\cite{New_Pen}).
\section{Equations of motion}%
\label{EqMo}%
\subsection{Euler-Lagrange equations.}%
The Lagrangian of the twistor variational principle derived in the
previous section has the form
\begin{equation}
{\cal L}%
= \varepsilon^{\mu\nu}\partial_\mu Z^\alpha\overline{Z}_\alpha
\Bigl(\overline{Z}_\beta\partial_\nu W^\beta +
\overline{W}_\beta\partial_\nu Z^\beta \Bigr),%
\label{Lagrangian}
\end{equation}
where the indices $\mu$, $ \nu$ run over $\tau$, $\sigma$ and
$\varepsilon^{\tau \sigma}=-\varepsilon^{\sigma\tau}=1$. We also
write $\partial_\mu = \partial/\partial\xi^\mu =
(\partial/\partial\tau,\,\partial/\partial\sigma)$ and extensively
use the shorthand notations ``\hspace{.5ex}$\dot{}$\hspace{.5ex}''
= $\partial_\tau$ and ``\hspace{.5ex}$\acute{}$\hspace{1ex}'' =
$\partial_\sigma$. The Euler-Lagrange equations are
\begin{equation}
\frac{\partial {\cal L}}{\partial \Upsilon^u} -
\frac{\partial}{\partial\xi^\mu} \left[\frac{\partial {\cal
L}}{\partial(\partial_\mu \Upsilon^u)}\right]=0.
\end{equation}
Here $\Upsilon^u$ = $\{Z^\alpha, W^\alpha, \overline{Z}_\alpha,
\overline{W}_\alpha\}$ are the dynamical quantities. The
substitution of the Lagrangian density (\ref{Lagrangian}) into the
equations yields
\begin{widetext}
\begin{eqnarray}
& \varepsilon^{\mu\nu}\left(\overline{Z}_\alpha\partial_\mu
\overline{Z}_\beta
\partial_\nu W^\beta + \partial_\mu\overline{Z}_\alpha\overline{Z}_\beta
\partial_\nu W^\beta + \overline{Z}_\alpha\partial_\mu\overline{W}_\beta
\partial_\nu Z^\beta + \partial_\mu\overline{Z}_\alpha\overline{W}_\beta
\partial_\nu Z^\beta \hspace{-0.25mm}-
\overline{W}_\alpha\partial_\mu\overline{Z}_\beta\partial_\nu
Z^\beta \hspace{-0.25mm}-
\partial_\mu\overline{W}_\alpha\overline{Z}_\beta\partial_\nu Z^\beta\right)
= 0, & \nonumber \\
& \varepsilon^{\mu\nu}\left(\partial_\mu
Z^\alpha\overline{Z}_\beta
\partial_\nu W^\beta + \partial_\mu
Z^\alpha\overline{W}_\beta\partial_\nu Z^\beta -
\partial_\mu W^\alpha\overline{Z}_\beta\partial_\nu Z^\beta\right) =
0,& \label{MotionEqns} \\
& \varepsilon^{\mu\nu}
\left(\overline{Z}_\alpha\partial_\mu\overline{Z}_\beta +
\partial_\mu\overline{Z}_\alpha
\overline{Z}_\beta\right)\partial_\nu Z^\beta = 0, \hspace{1.4ex}
\varepsilon^{\mu\nu}\partial_\mu Z^\alpha\overline{Z}_\beta
\partial_\nu Z^\beta = 0. & \nonumber
\end{eqnarray}
\end{widetext}

We can rewrite these equations in terms of the spinor fields
$\bar{\pi}^A$ and $\bar{\eta}^A$ and the world-sheet derivatives
of $x^{AA'}$ employing the definitions of the null twistors
$Z^\alpha$ and $W^\alpha$. In what follows it will be also
convenient to take the advantage of the identities
\begin{eqnarray}
\bar{\pi}^A \pi^{A'}\bar{\eta}^B \hspace{-1ex}& - &\hspace{-1ex}
(\bar{\pi}^A\eta^{A'} + \pi^{A'}\bar{\eta}^A)\bar{\pi}^B =
\pi^{A'}\epsilon^{AB} -
\eta^{A'}\bar{\pi}^A\bar{\pi}^B, \nonumber \\
\dot{x}^{AA'}\acute{x}^{BB'} \hspace{-1ex}& - &\hspace{-1ex}
\acute{x}^{AA'}\dot{x}^{BB'} = \phi^{AB}\epsilon^{A'B'} +
\bar{\phi}^{A'B'}\epsilon^{AB}.%
\label{Identities}
\end{eqnarray}
Here we use the normalization condition
\begin{equation}
\varepsilon_{AB} = \bar{\pi}_A\bar{\eta}_B -
\bar{\eta}_A\bar{\pi}_B%
\label{SymForm}
\end{equation}
and the symmetric spin-tensor $\phi^{AB}$ is given by the
expression $\phi^{AB}$ = $\dot{x}^{(A}{}_{C'}\acute{x}^{B)\,C'}$.

Let us consider the first equation in the system
(\ref{MotionEqns}). Utilizing the formulae presented above, we
obtain that this equation is equivalent to the following system:
\begin{widetext}
\begin{eqnarray}
\dot{x}{}^{BB'}[x^{AA'}(\pi_{B'}\epsilon_{AB} +
\eta_{B'}\bar{\pi}_A\bar{\pi}_B)]\,\acute{} & - &
\acute{x}{}^{BB'}[x^{AA'}(\pi_{B'}\epsilon_{AB} +
\eta_{B'}\bar{\pi}_A\bar{\pi}_B)]\,\dot{} = 0, \nonumber \\
\dot{x}^{BB'}\left(\pi_{B'}\epsilon_{AB} +
\eta_{B'}\bar{\pi}_A\bar{\pi}_B\right)\acute{} & - &
\acute{x}\left(\pi_{B'}\epsilon_{AB} +
\eta_{B'}\bar{\pi}_A\bar{\pi}_B\right)\dot{} = 0. \label{EM1.p}
\end{eqnarray}
\end{widetext}
Substituting the second equation in (\ref{EM1.p}) into the first
one and using the second identity in (\ref{Identities}), we can
represent the first equation in (\ref{MotionEqns}) in the form
\begin{widetext}
\begin{eqnarray}
\dot{x}^{BB'}[\bar{\eta}_{A}\bar{\pi}_B\pi_{B'} -
\bar{\pi}_A(\bar{\pi}_B\eta_{B'} +
\bar{\eta}_B\pi_{B'})]\,\acute{} & - &
\acute{x}^{BB'}[\bar{\eta}_{A}\bar{\pi}_B\pi_{B'} -
\bar{\pi}_A(\bar{\pi}_B\eta_{B'} + \bar{\eta}_B\pi_{B'})]\,\dot{}
= 0, \nonumber \\ 2\bar{\phi}{}^{A'B'}\pi_{B'} & + &
\eta^{A'}(\phi^{AB}\bar{\pi}_A\bar{\pi}_{B})  = 0. \label{EM1}
\end{eqnarray}
\end{widetext}
The second equation in the system (\ref{MotionEqns}) yields
\begin{widetext}
\begin{eqnarray}
\lefteqn{\dot{x}{}^{BB'}[(x^{AA'}\eta_{A'})\,\acute{}\,\bar{\pi}_B\pi_{B'}
- (x^{AA'}\pi_{A'})\,\acute{}\,(\bar{\pi}_B\eta_{B'} +
\bar{\eta}_B\pi_{B'})] - } \hspace{5.5cm} \nonumber \\ & &
\hspace{-3.1cm}
\acute{x}{}^{BB'}[(x^{AA'}\eta_{A'})\,\dot{}\,\bar{\pi}_B\pi_{B'}
- (x^{AA'}\pi_{A'})\,\dot{}\,(\bar{\pi}_B\eta_{B'} +
\bar{\eta}_B\pi_{B'})] = 0, \nonumber \\
\lefteqn{\dot{x}{}^{BB'}[\acute{\eta}_{A'}\bar{\pi}_B\pi_{B'} -
\acute{\pi}_{A'}(\bar{\pi}_B\eta_{B'} + \bar{\eta}_B\pi_{B'})] - }
\hspace{5.5cm} \nonumber \\ & &
\hspace{-5.5mm}\acute{x}{}^{BB'}[\dot{\eta}_{A'}\bar{\pi}_B\pi_{B'}
- \dot{\pi}_{A'}(\bar{\pi}_B\eta_{B'} + \bar{\eta}_B\pi_{B'})] =
0.
\end{eqnarray}
\end{widetext}
Substituting the second equation in this system into the first and
using (\ref{Identities}) again, we find that the second equation
in (\ref{MotionEqns}) results in
\begin{widetext}
\begin{eqnarray}
\dot{x}{}^{BB'}[\acute{\eta}_{A'}\bar{\pi}_B\pi_{B'} -
\acute{\pi}_{A'}(\bar{\pi}_B\eta_{B'} + \bar{\eta}_B\pi_{B'})] & -
& \acute{x}{}^{BB'}[\dot{\eta}_{A'}\bar{\pi}_B\pi_{B'} -
\dot{\pi}_{A'}(\bar{\pi}_B\eta_{B'} + \bar{\eta}_B\pi_{B'})] = 0,
\nonumber \\ 2\phi^{AB}\bar{\pi}_{B} & + &
\bar{\eta}^{A}(\bar{\phi}^{A'B'}\pi_{A'}\pi_{B'})  = 0.
\label{EM2}
\end{eqnarray}
\end{widetext}
We incidentally observe that the second equations in the systems
(\ref{EM1}) and (\ref{EM2}) are complex conjugates of one another.
We next consider the third equation of the system
(\ref{MotionEqns}). It can be presented as follows:
\begin{eqnarray}
\dot{x}{}^{BB'}(x^{AA'}\bar{\pi}_A\bar{\pi}_B\pi_{B'})\,\acute{} &
- &
\acute{x}{}^{BB'}(x^{AA'}\bar{\pi}_A\bar{\pi}_B\pi_{B'})\,\dot{} =
0, \nonumber \\
\dot{x}{}^{BB'}(\bar{\pi}_A\bar{\pi}_B\pi_{B'})\,\acute{} & - &
\acute{x}{}^{BB'}(\bar{\pi}_A\bar{\pi}_B\pi_{B'})\,\dot{} = 0.
\end{eqnarray}
The substitution of the second equation above into the first and
the use of (\ref{Identities}) allow us to write this pair of
equations in the form
\begin{eqnarray}
\dot{x}{}^{BB'}(\bar{\pi}_A\bar{\pi}_B\pi_{B'})\,\acute{} -
\acute{x}{}^{BB'}(\bar{\pi}_A\bar{\pi}_B\pi_{B'})\,\dot{} & = & 0,
\nonumber \\ \phi^{AB}\bar{\pi}_A\bar{\pi}_B & = & 0. \label{EM3}
\end{eqnarray}
The fourth equation in (\ref{MotionEqns}) gives
\begin{eqnarray}
[\dot{x}{}^{BB'}(x^{AA'}\pi_{A'})\,\acute{} & - &
\acute{x}{}^{BB'}(x^{AA'}\pi_{A'})\,\dot{}\,]\bar{\pi}_B\pi_{B'} =
0, \nonumber \\ (\dot{x}{}^{BB'}\acute{\pi}_{A'} & - &
\acute{x}{}^{BB'}\dot{\pi}_{A'})\bar{\pi}_B\pi_{B'} = 0.
\label{SM}
\end{eqnarray}
The second equation in~(\ref{SM}) can be used to rewrite these two
equations as follows:
\begin{eqnarray}
(\dot{x}{}^{BB'}\acute{\pi}_{A'} -
\acute{x}{}^{BB'}\dot{\pi}_{A'})\bar{\pi}_B\pi_{B'} & = & 0,
\nonumber \\ \bar{\phi}{}^{A'B'}\pi_{A'}\pi_{B'} & = & 0.
\label{EM4}
\end{eqnarray}
We also find that the second equation in (\ref{EM3}) is complex
conjugate of the second equation in~(\ref{EM4}).

Let us use the first equation in (\ref{EM2}) to simplify the first
equation in the system (\ref{EM1}). The calculation gives
\begin{eqnarray}
& &%
\hspace{-1em}[\dot{x}{}^{BB'}(\bar{\pi}_B\eta_{B'} +
\bar{\eta}_B\pi_{B'})\,\acute{} -
\acute{x}{}^{BB'}(\bar{\pi}_B\eta_{B'} +
\bar{\eta}_B\pi_{B'})\,\dot{}\,]\bar{\pi}_A -
\nonumber \\
& &%
\hspace{-1em}[\dot{x}{}^{BB'}(\bar{\pi}_B\pi_{B'})\,\acute{} -
\acute{x}{}^{BB'}(\bar{\pi}_B\pi_{B'})\,\dot{}\,]%
\bar{\eta}_A = 0.
\end{eqnarray}
Since $\bar{\pi}{}^A$ and $\bar{\eta}{}^A$ con\-sti\-tute a
nor\-mal\-ized ba\-sis for the two-dimensional vector space
$\mathbb{C}^2$, the equation above is equivalent to the following
pair:
\begin{eqnarray}
\dot{x}^{AA'}(\bar{\pi}_A\eta_{A'} +
\bar{\eta}_A\pi_{A'})\,\acute{} & - &
\acute{x}^{AA'}(\bar{\pi}_A\eta_{A'} +
\bar{\eta}_A\pi_{A'})\,\dot{} = 0, \nonumber \\
\dot{x}^{AA'}(\bar{\pi}_A\pi_{A'})\,\acute{} & - &
\acute{x}^{AA'}(\bar{\pi}_A\pi_{A'})\,\dot{} = 0. \label{REM}
\end{eqnarray}
Performing the same procedure with the first equations in
(\ref{EM3}) and (\ref{EM4}), we derive the equation
\begin{equation}
\dot{x}^{AA'}(\bar{\pi}_A\pi_{A'})\,\acute{} -
\acute{x}^{AA'}(\bar{\pi}_A\pi_{A'})\,\dot{} = 0.
\end{equation}
It coincides with the second equation in the system (\ref{REM}).

Finally, for the purposes of the future analysis, we divide the
independent Euler-Lag\-ran\-ge equations in the following three
pairs:
\begin{eqnarray}
& &%
(\dot{x}^{AA'}\acute{\bar{\pi}}_{B} -
\acute{x}^{AA'}\dot{\bar{\pi}}_{B}) (\bar{\pi}_A\eta_{A'} +
\pi_{A'}\bar{\eta}_A) - \nonumber \\
& &%
\hspace{4.7em}(\dot{x}^{AA'}\acute{\bar{\eta}}_{B} -
\acute{x}^{AA'}\dot{\bar{\eta}}_B)\bar{\pi}_A\pi_{A'} = 0,%
\label{Pair1} \\%
& &%
\dot{x}^{AA'}\left(\bar{\pi}_A\eta_{A'} +
\pi_{A'}\bar{\eta}_A\right)\acute{} -
\acute{x}^{AA'}\left(\bar{\pi}_A\eta_{A'} +
\pi_{A'}\bar{\eta}_A\right)\dot{} = 0; \nonumber
\end{eqnarray}
\begin{eqnarray}
\dot{x}^{AA'}(\bar{\pi}_A\pi_{A'})\,\acute{} -
\acute{x}^{AA'}(\bar{\pi}_A\pi_{A'})\,\dot{} & = & 0, \nonumber \\
(\dot{x}^{AA'}\acute{\bar{\pi}}_{B} -
\acute{x}^{AA'}\dot{\bar{\pi}}_{B})\bar{\pi}_A\pi_{A'} & = & 0;
\label{Pair2}
\end{eqnarray}
\begin{eqnarray}
2\bar{\phi}^{A'B'}\pi_{A'} +
(\phi^{AB}\bar{\pi}_A\bar{\pi}_B)\eta^{B'} & = & 0, \nonumber \\
\phi^{AB}\bar{\pi}_A\bar{\pi}_B & = & 0. \label{Pair3}
\end{eqnarray}
Here we made the substitutions of dummy indices where appropriate
and presented complex conjugate versions of some equations.
\subsection{Preliminary analysis}%
\hspace{5mm}%
Now we shall establish a few auxiliary results.

The third pair of the motion equations (\ref{Pair3}) gives
$\phi^{AB}\bar{\pi}_B = 0$, which means that
\begin{equation}
2\phi^{AB} = v\bar{\pi}^A\bar{\pi}^B \label{Null_st}
\end{equation}
for some complex-valued function $v(\tau,\sigma)$.

Let us substitute the representation (\ref{Null_st}) into the
second equation of (\ref{Identities})
\begin{equation}
2(\dot{x}^{AA'}\acute{x}^{BB'} - \acute{x}^{AA'}\dot{x}^{BB'}) =
v\bar{\pi}^A\bar{\pi}^B\epsilon^{A'B'} +
\bar{v}\pi^{A'}\pi^{B'}\epsilon^{AB}.
\end{equation}
Multiplying both sides of this equation by $\dot{x}_{BB'}$ we
obtain
\begin{equation}
2[\dot{x}^2\acute{x}^{AA'} - (\dot{x}\acute{x})\dot{x}^{AA'}] =
v\bar{\pi}^A\bar{\pi}_B\dot{x}^{BA'} +
\bar{v}\pi^{A'}\pi_{B'}\dot{x}^{AB'}. \label{Technical_4.3}
\end{equation}

Following Ref.~\cite{Gusev-Zheltukhin} we calculate
\begin{eqnarray}
2\dot{x}^A{}_{C'}\acute{x}^{BC'} & = &
2\dot{x}^{(A}{}_{C'}\acute{x}^{B)\,C'} +
      2\dot{x}^{[A}{}_{C'}\acute{x}^{B]\,C'}
\nonumber \\ & = & 2\phi^{AB} + (\dot{x}\acute{x})\epsilon^{AB}
\nonumber \\ & = & v\bar{\pi}^A\bar{\pi}^B +
(\dot{x}\acute{x})\epsilon^{AB}.
\end{eqnarray}
Using this result one obtains
\begin{equation}
2\dot{x}^A{}_{C'}\acute{x}^{BC'}\dot{x}_{AD'}\acute{x}_B{}^{D'} =
(\dot{x}\acute{x})^2.%
\label{Technical_4.1}
\end{equation}
On the other hand,
\begin{equation}
\dot{x}^A{}_{D'}\acute{x}^{BC'}\dot{x}_{AC'}\acute{x}_B{}^{D'} =
-\dot{x}^A{}_{C'}\dot{x}_{AD'}\acute{x}^{BC'}\acute{x}_B{}^{D'}
\end{equation}
and this fact can be used to show that the left hand side of
equation (\ref{Technical_4.1}) is also given by
\begin{eqnarray}
2\dot{x}^A{}_{C'}\acute{x}^{BC'}\dot{x}_{AD'}\acute{x}_B{}^{D'} =
& & \nonumber \\%
& & \hspace{-7.5em} = (\dot{x}^A{}_{C'}\dot{x}_{AD'} -
\dot{x}^A{}_{D'}\dot{x}_{AC'})
      \acute{x}^{BC'}\acute{x}_B{}^{D'} \nonumber \\
& & \hspace{-7.5em} =
2\dot{x}^A{}_{[C'}\dot{x}_{|A|\,D']}\acute{x}^{BC'}\acute{x}_B{}^{D'}
  = \dot{x}^2\acute{x}^2.
\label{Technical_4.2}
\end{eqnarray}
The equations (\ref{Technical_4.1}) and (\ref{Technical_4.2})
result in the identity (\ref{NullDet}).
The left hand side of (\ref{NullDet}) is the determinant of the
induced metric on the null string world-sheet, or equivalently, on
a two-dimensional real null submanifold of 4D~Minkowski
space-time, and this equation shows that it vanishes identically.
The vanishing property of the determinant of the induced metric is
invariant under the group of non-degenerate diffeomorphisims of
the null string world-sheet
\begin{equation}
\tau \rightarrow f(\tau,\sigma), \,\,\,\, \sigma \rightarrow
\varphi(\tau,\sigma). \label{GDiff}
\end{equation}

According to the second Noether
theorem~\cite{Noether,Barbashov-Nesterenko}, the reparametrization
invariance (\ref{GDiff}) of the twistor action functional implies
that the equations of motion contain two arbitrary real-valued
functions. Then, without loss of generality, we can choose one of
them in such a way as to ensure that $\dot{x}^2 = 0$. Taking into
account~(\ref{NullDet}), this entails
\begin{equation}
\dot{x}^2 = 0,\,\,\,\,\dot{x}\acute{x} = 0.%
\label{OGauge}
\end{equation}
Having fixed the orthogonal gauge~(\ref{OGauge}), we restrict the
group of diffeomorphisms of the null string world-sheet to the
following subgroup of transformations:
\begin{equation}
\tau \rightarrow f(\tau,\sigma), \,\,\,\, \sigma \rightarrow
\varphi(\sigma). \label{RDiff}
\end{equation}
It, therefore, follows from~(\ref{Technical_4.3}) and
(\ref{OGauge}) that
\begin{equation}
v\bar{\pi}^A\dot{x}^{BA'}\bar{\pi}_B +
\bar{v}\pi_{B'}\dot{x}^{AB'}\pi^{A'} = 0,
\end{equation}
and projection of this equation on the elements of the spin-tensor
basis $\bar{\pi}_A\pi_{A'}$, $\bar{\pi}_A\eta_{A'}$,
$\pi_{A'}\bar{\eta}_{A'}$ and $\bar{\eta}_A\eta_{A'}$ yields
\begin{eqnarray}
v\dot{x}^{AA'}\bar{\pi}_A\eta_{A'} +
\bar{v}\dot{x}^{AA'}\bar{\eta}_A\pi_{A'} = 0, \nonumber \\
\dot{x}^{AA'}\bar{\pi}_A\pi_{A'} = 0.%
\label{Technical_4.4}
\end{eqnarray}
The second equation in the system~(\ref{Technical_4.4}), together
with the null property~(\ref{OGauge}) of the vector field
$\dot{x}^a$, gives rise to the representation for this vector
field in the form
\begin{equation}
\dot{x}^{AA'} = r\bar{\pi}^A\pi^{A'}, \label{DotX}
\end{equation}
where $r(\tau,\sigma)$ is a real-valued function. This
representation for one of the two vector fields tangent to the
null string world-sheet automatically solves the first equation of
the system (\ref{Technical_4.4}). The result~(\ref{DotX}) and the
second equation in (\ref{OGauge}) imply
\begin{equation}
\acute{x}^{AA'}\bar{\pi}_A\pi_{A'}= 0.
\end{equation}
Now $\acute{x}{}^{AA'}$ can be written in the form
\begin{equation}
\acute{x}^{AA'} = \bar{\zeta}\bar{\pi}^A\eta^{A'} +
\zeta\bar{\eta}^A\pi^{A'} +  g\bar{\pi}^A\pi^{A'},
\end{equation}
where $\zeta(\tau,\sigma)$ and $g(\tau,\sigma)$ are complex- and
real-valued func\-tions, re\-spec\-ti\-ve\-ly. The\-re\-fore, the
representation for two linearly independent vector fields tangent
to the null string world-sheet which obey constraints
(\ref{OGauge}) has the form \begin{equation} \dot{x}^{AA'} =
r\bar{\pi}^A\pi^{A'},\,\,\,\, \acute{x}^{AA'} =
\bar{\zeta}\bar{\pi}^A\eta^{A'} +
                  \zeta\bar{\eta}^A\pi^{A'} + g\bar{\pi}{}^A\pi^{A'}.
\label{TanVecs}
\end{equation}
The substitution of the representation (\ref{TanVecs}) into the
first equation of the system (\ref{Pair1}) gives
\begin{equation}
\dot{\bar{\pi}}{}^A(\zeta + \bar{\zeta}) = 0.%
\label{Khlopok}
\end{equation}

The situation when $\dot{\bar{\pi}}{}^A$ vanishes corresponds to
the geodesic property of the null string world-sheets and it has
been considered in Ref.~\cite{Gusev-Zheltukhin}. Unfortunately,
the authors of that article had overlooked the other possibility,
given by the vanishing of the expression in the round brackets in
the equation~(\ref{Khlopok}), and had not paid any attention to
non-geodesic null strings. The remainder of this paper will be
devoted to the analysis of the non-geodesic case, which
corresponds, as we shall see, to the situation when the function
$\zeta (\tau,\sigma)$ is purely imaginary.

It is convenient to redefine the function $\zeta$ to be a
real-valued function in the null string motion equations and in
the representations for the vector fields $\dot{x}{}^{AA'}$ and
$\acute{x}{}^{AA'}$ by means of the substitution $\zeta$
$\rightarrow$ $-i\zeta$. Taking into account (\ref{TanVecs}), we
can reduce the remainder of the motion equations
(\ref{Pair1})~--~(\ref{Pair2}) to the system
\begin{eqnarray}
r(\bar{\pi}^A\acute{\bar{\pi}}_A + \pi^{A'}\acute{\pi}_{A'}) & = &
\nonumber \\%
2i\zeta(\pi^{A'}\dot{\eta}_{A'} - \bar{\pi}^A\dot{\bar{\eta}}_A) &
+ & g(\bar{\pi}^A\dot{\bar{\pi}}_A +
\pi^{A'}\dot{\pi}_{A'}), \\%
\bar{\pi}^A\dot{\bar{\pi}}_A - \pi^{A'}\dot{\pi}_{A'} & = & 0.
\nonumber
\end{eqnarray}
Finally, we obtain the motion equations of the null string in the
form
\begin{eqnarray}
\dot{x}^{AA'} = r\bar{\pi}^A\pi^{A'}, \acute{x}^{AA'}\!\!&=&\!\!
\hphantom{2}i\zeta (\bar{\pi}{}^A\eta^{A'} -
\bar{\eta}{}^A\pi^{A'}) +
g\bar{\pi}{}^A\pi^{A'}, \nonumber \\
r(\bar{\pi}^A\acute{\bar{\pi}}_A + \pi^{A'}\acute{\pi}_{A'}) &=&
2i\zeta(\pi^{A'}\dot{\eta}_{A'} - \bar{\pi}^A\dot{\bar{\eta}}_A) +
\label{MEqns} \\%
g(\bar{\pi}{}^A\dot{\bar{\pi}}_A + \pi^{A'}\dot{\pi}_{A'}), &&
\bar{\pi}^A\dot{\bar{\pi}}_A - \pi^{A'}\dot{\pi}_{A'} = 0,
\nonumber%
\end{eqnarray}
where $r(\tau,\sigma)$ and $\zeta(\tau,\sigma)$ are arbitrary
real-valued functions.

Let us briefly explore the effects of the gauge transformations
(\ref{Scale_Add}) and the null string world-sheet
reparametrizations (\ref{RDiff}) on the equations of motion. The
gauge transformations (\ref{Scale_Add}) leave the null string
motion equations (\ref{MEqns}) invariant and result in simple
redefinitions of the functions $r$ and $g$
\begin{equation}
r \rightarrow q^2 r, \,\,\,\, g \rightarrow q^2 g - iq(p -
\bar{p})\zeta.
\end{equation}
Under the world-sheet reparametrizations (\ref{RDiff})
\begin{equation}
\dot{x}{}^a  \rightarrow \dot{f}{}^{-1}\dot{x}{}^a,
\,\,\,\,\,\,\,\, \acute{x}^a  \rightarrow
\acute{\varphi}{}^{-1}\acute{x}{}^a -
(\acute{\varphi}\dot{f}){}^{-1}\acute{f}\dot{x}{}^a.
\end{equation}
We then observe that the reparametrizations (\ref{RDiff}) preserve
the form of the null string motion equations (\ref{MEqns}) while
leading to the following redifinitions of the functions $r$, $g$
and $\zeta$:
\begin{equation}
r \rightarrow \dot{f}{}^{-1}r, \,\, g \rightarrow
\acute{\varphi}{}^{-1}g -
              (\acute{\varphi}\dot{f}){}^{-1}\acute{f}r, \,\,
\zeta \rightarrow \acute{\varphi}{}^{-1}\zeta.
\end{equation}
One can easily check that the null property of the vector field
$\dot{x}{}^a$ and the orthogonal character of the vector fields
$\dot{x}{}^a$ and $\acute{x}^a$ are preserved with respect to both
those transformations.

The invariant property of the twistor action functional with
respect to either of those transformations can be used in order to
eliminate the null component, $g\bar{\pi}{}^A\pi^{A'}$, of the
space-like vector field $\acute{x}{}^{AA'}$ from the null string
equations of motion (\ref{MEqns}). This can be achieved by
performing the transformations (\ref{Scale_Add}) with the
parameters
\begin{equation}
q = 1, \,\,\,\, i(p-\bar{p}) = \zeta^{-1}g,
\end{equation}
where the real part of the function $p$ may remain arbitrary.
Then, the null string equations of motion take the reduced form
\begin{eqnarray}
\dot{x}^{AA'} = r\bar{\pi}^A\pi^{A'},\,\,\,\, \acute{x}^{AA'} & =
& i\zeta (\bar{\pi}{}^A\eta^{A'} - \bar{\eta}{}^A\pi^{A'}),
\nonumber \\ r(\bar{\pi}^A\acute{\bar{\pi}}_A +
\pi^{A'}\acute{\pi}_{A'}) & = & 2i\zeta(\pi^{A'}\dot{\eta}_{A'} -
\bar{\pi}^A\dot{\bar{\eta}}_A), \nonumber \\
\bar{\pi}^A\dot{\bar{\pi}}_A - \pi^{A'}\dot{\pi}_{A'} & = & 0.
\label{RMEqns}
\end{eqnarray}
They are invariant under the gauge transformations
(\ref{Scale_Add}) with real functions $p(\tau,\sigma)$. These
restricted gauge transformations result in trivial rescaling of
the function $r(\tau,\sigma)$ in the null string motion equations
(\ref{RMEqns})
\begin{equation}
r \rightarrow q^2 r
\end{equation}
and reflect the freedom inherent in the choice of the extent of
the null direction represented by the vector field $\dot{x}{}^a$.
This restriction of the admissible gauge transformations to those
with real $p$s further reduces the reparametrization freedom of
the null string world-sheet. The invariance of the null string
equations of motion in the form (\ref{RMEqns}) requires that the
function $f$ entering the reparametrization transformations
(\ref{RDiff}) is a function of $\tau$ alone, thereby restricting
the reparametrization freedom to the following transformations:
\begin{equation}
\tau \rightarrow f(\tau), \,\,\,\, \sigma \rightarrow \varphi
(\sigma).%
\label{TDiff}
\end{equation}

In the equations above the function $r$ defines the extent of the
flagpole direction $l^a\equiv\bar{\pi}^A\pi^{A'}$ tangent to the
null string world-sheet. We also note that unrestricted gauge
transformations~(\ref{Scale_Add}) preserve the associated flag
plane represented by the space-like vector field
$\mu^a\equiv\bar{\pi}^A\eta^{A'}+\pi^{A'}\bar{\eta}^A$. This
vector is orthogonal to the space-like vector field $q^a\equiv
i(\bar{\pi}{}^A\eta^{A'} - \bar{\eta}{}^A\pi^{A'})$ tangent to the
null string world-sheet. The vector fields $l^a$, $\mu^a$, and
$q^a$ together with the second null vector field
$n^a\equiv\bar{\eta}{}^A\eta^{A'}$ define a (non-normalized)
Newman-Penrose tetrad for 4D~Minkowski space-time. Here the vector
fields $\mu^a$ and $q^a$ can be expressed in the terms of the
usual complex elements of the tetrad as follows:
\begin{equation}
\mu^a = m^a + \bar{m}{}^a, \,\,\,\, q^a = i(m^a - \bar{m}{}^a).
\end{equation}
\subsection{Integrability conditions}%
\label{ComConds}%
\hspace{5mm}%
The representations for $\dot{x}{}^{AA'}$ and $\acute{x}{}^{AA'}$
in (\ref{RMEqns}) must satisfy the compatibility conditions
\begin{equation} (\dot{x}{}^{AA'})\acute{} = (\acute{x}{}^{AA'})\dot{}.
\label{ComCon}
\end{equation}
In turn, this leads to some compatibility conditions on the
$\tau$-~and $\sigma$-derivatives of the basis spinor fields
$\bar{\pi}{}^A$ and $\bar{\eta}{}^A$.  It will have proven
interesting to explore the geometrical significance of the
compatibility conditions~(\ref{ComCon}).

Using the definitions of the vector fields $l^a$ and $q^a$ we find
\begin{eqnarray}
\frac{\partial}{\partial\tau} = \frac{\partial
x^a}{\partial\tau\hspace{2mm}} \frac{\partial}{\partial x^a} &
\equiv &
\dot{x}{}^a \nabla_a = rl^a \nabla_a, \nonumber \\
\frac{\partial}{\partial\sigma} = \frac{\partial
x^a}{\partial\sigma\hspace{2mm}} \frac{\partial}{\partial x^a} &
\equiv & \acute{x}{}^a \nabla_a =\zeta q^a \nabla_a.
\label{Technical_3.4.0}
\end{eqnarray}
We next calculate the Lie derivative of the vector field
$\partial_\sigma$ along $\partial_\tau$
\begin{equation}
\pounds_{rl}(\zeta q^a) = \tilde{a}l^a + \tilde{b}q^a
                             + r\zeta\pounds_l\eta^a,
\end{equation}
where we have defined $\tilde{a} = -\zeta q^a\nabla_ar$ and
$\tilde{b} = rl^a\nabla_a\zeta$. In the derivation we have used
the fact that in 4D~Minkowski space-time the derivatives
$\nabla_a$ commute. Now, geometrical meaning of the
condition~(\ref{ComCon}) becomes apparent, it requires the Lie
derivative of the connecting vector field $q^a$ along the vector
field $l^a$ to be contained in the subspace spanned by those
vector fields
\begin{equation}
\pounds_l\eta^a = al^a + bq^a. \label{FrobInteg}
\end{equation}
Here $a = -r^{-1}q^a\nabla_a r$ and $b =
\zeta^{-1}l^a\nabla_a\zeta$ are some real-valued functions of
$\tau$ and $\sigma$. This equation can be recognized as the
Frobenius integrability condition applied to the vector fields
$l^a$ and $q^a$. We note in passing that the vector field $\zeta
q^a$ plays the role of the Jacobi field along the null congruence
given by the vector field $rl^a$ and, therefore, is simply a
Lie-dragged vector field. This condition can be phrased in a yet
another form by observing that (\ref{FrobInteg}) entails that the
projections of the vector field $\pounds_l\eta^a$ on the elements
$l^a$ and $\mu^a$ of the Newman-Penrose tetrad must vanish
\begin{equation}
l_a\pounds_l q^a = 0, \,\,\,\, \mu_a\pounds_l q^a = 0.
\label{FrobProj}
\end{equation}

Having established the geometrical meaning of the compatibility
conditions (\ref{ComCon}), or equivalently~(\ref{FrobProj}), we
can proceed with their analysis. The first equation
in~(\ref{FrobProj}) gives
\begin{equation}
l^al^b\nabla_b q_a - l^a q^b\nabla_b l_a = 0.
\end{equation}
Noting that the second term on the left hand side vanishes
identically and using the orthogonal property of the vector fields
$l^a$ and $q^a$ we obtain the equation
\begin{equation}
q^a l^b\nabla_b l_a = 0%
\label{FrobPair1}
\end{equation}
as the first integrability condition. The second equation
in~(\ref{FrobProj}) yields
\begin{equation}
\mu^a q^b\nabla_b l_a - \mu^a l^b\nabla_b q_a = 0.%
\label{FrobPair2}
\end{equation}
Summarizing, the Frobenius integrability conditions for the
two-dimensional submanifold of 4D~Minkowski space-time
representing the null string world-sheet are given by the formulae
(\ref{FrobPair1}) and (\ref{FrobPair2}).
A straightforward calculation shows that these equations are
invariant under the restricted (to real $q$s and $p$s) gauge
transformations of the form~(\ref{Scale_Add}).

At this point we can compare the different forms of the
integrability conditions, namely, the differential condition for
the null bivector $p_{ab}(x)$ stated in Sec.~\ref{Intro} and
(\ref{FrobInteg}). For this purpose, we note that the null
bivector can be written in the form
\begin{equation}
2p_{ab} = \bar{\pi}{}_A\bar{\pi}{}_B\epsilon_{A'B'} +
          \pi_{A'}\pi_{B'}\epsilon_{AB}.%
\label{DiffCon}
\end{equation}
Here we have used the definitions (\ref{P_ab}) and (\ref{Normals})
for the null bivector and the vector fields $u^a$ and $v^a$, the
normalization conditions (\ref{U_VNorm}) and identity
(\ref{SymForm}). The integrability condition mentioned above is
equivalent to the following equation:
\begin{equation}
p_{ab}\nabla_c p_{de}\epsilon^{bcde} = 0,%
\label{Main}
\end{equation}
where the totally antisymmetric tensor density $\epsilon^{bcde}$
is given by the expression, \cite[Vol. 1]{SST},
\begin{equation}
\epsilon^{bcde} =
i(\epsilon^{BD}\epsilon^{CE}\epsilon^{B'E'}\epsilon^{C'D'}
                  -\epsilon^{BE}\epsilon^{CD}\epsilon^{B'D'}\epsilon^{C'E'}).
\end{equation}
First, we write
\begin{eqnarray}
2\nabla_c p_{de} & = &
        \bar{\pi}_E\epsilon_{D'E'}\nabla_{CC'}\bar{\pi}_D +
        \bar{\pi}_D\epsilon_{D'E'}\nabla_{CC'}\bar{\pi}_E +
\nonumber \\%
&&\hspace{-2em}\pi_{E'}\epsilon_{DE}\nabla_{CC'}\pi_{D'} +
        \pi_{D'}\epsilon_{DE}\nabla_{CC'}\pi_{E'}.
\end{eqnarray}
Second, we calculate
\begin{eqnarray}
\epsilon^{bcde}\nabla_c p_{de} & = &
i(\bar{\pi}{}^B\nabla^{CB'}\bar{\pi}_C -
\pi^{B'}\nabla^{BC'}\pi_{C'} + \nonumber \\%
&& \hspace{.75em}\bar{\pi}{}_C\nabla^{CB'}\bar{\pi}^B -
\pi_{C'}\nabla^{BC'}\pi^{B'}).
\end{eqnarray}
Finally, we obtain
\begin{widetext}
\begin{eqnarray}
2p_{ab}\nabla_c p_{de}\epsilon^{bcde} & = &
i[\bar{\pi}_A\bar{\pi}^B\pi^{B'}\nabla_{BB'}\pi_{A'}
     - \pi_{A'}\bar{\pi}{}^B\pi^{B'}\nabla_{BB'}\bar{\pi}_A +
\pi_{A'}\pi^{B'}\pi^{C'}\nabla_{AC'}\pi_{B'} - \nonumber \\
&& \bar{\pi}_A\bar{\pi}{}^B\bar{\pi}^C\nabla_{CA'}\bar{\pi}_B +
\bar{\pi}_A\pi_{A'}(\bar{\pi}{}^B\nabla_{BB'}\pi^{B'}
                           -\pi^{B'}\nabla_{BB'}\bar{\pi}_B)].
\label{Monster}
\end{eqnarray}
\end{widetext}
Then, the desired result follows from equating the right hand side
of the formula (\ref{Monster}) to zero. Since the
equation~(\ref{Main}) is equivalent to its projections on the
spin-tensor basis elements $\bar{\pi}{}^A\pi^{A'}$,
$\bar{\pi}{}^A\eta^{A'}$, $\bar{\eta}{}^A\pi^{A'}$ and
$\bar{\eta}^A\eta^{A'}$, we derive the results which are presented
below. Firstly, the projection of equation~(\ref{Main}) on
$\bar{\pi}{}^A\pi^{A'}$ vanishes identically. Secondly, its
projections on the spin-tensor basis elements
$\bar{\pi}{}^A\eta^{A'}$ and $\bar{\eta}^A\pi^{A'}$ are the
complex conjugates of one another and can be represented in the
form
\begin{equation}
\bar{\pi}{}^A\bar{\pi}^B\pi^{B'}\nabla_{BB'}\bar{\pi}_A -
\pi^{A'}\bar{\pi}^B\pi^{B'}\nabla_{BB'}\pi_{A'} = 0.
\label{Technical_3.4.1}
\end{equation}
Finally, the projection of~(\ref{Main}) on the remaining element
of the spin-tensor basis, $\bar{\eta}{}^A\eta^{A'}$, is given by
\begin{widetext}
\begin{eqnarray}
2(\bar{\pi}{}^A\bar{\pi}{}^B\pi^{B'}\nabla_{BB'}\bar{\eta}_A & - &
     \pi^{A'}\bar{\pi}{}^B\pi^{B'}\nabla_{BB'}\eta_{A'}) +
     \pi^{A'}\bar{\pi}{}^B\eta^{B'}\nabla_{BB'}\pi_{A'} -
\nonumber \\%
\pi^{A'}\bar{\eta}{}^B\pi^{B'}\nabla_{BB'}\pi_{A'}  & - &
     \bar{\pi}{}^A\bar{\eta}{}^B\pi^{B'}\nabla_{BB'}\bar{\pi}_A +
      \bar{\pi}{}^A\bar{\pi}{}^B\eta^{B'}\nabla_{BB'}\bar{\pi}_A  = 0.
\label{Technical_3.4.2}
\end{eqnarray}
\end{widetext}%
The substitution of the definitions for the vector fields $l^a$,
$q^a$ and $\mu^a$ through the spinors $\bar{\pi}^A$ and
$\bar{\eta}^A$ into (\ref{Technical_3.4.1}) and
(\ref{Technical_3.4.2}) reduces those equations to the system
(\ref{FrobPair1}) and (\ref{FrobPair2}). This concludes our
demonstration of the equivalence of the integrability conditions
(\ref{Main}) and (\ref{FrobInteg}).

Using (\ref{Technical_3.4.0}) we can rewrite the formulae
(\ref{Technical_3.4.1}) and (\ref{Technical_3.4.2}) as
\begin{eqnarray}
r(\bar{\pi}{}^A\acute{\bar{\pi}}_A + \pi^{A'}\acute{\pi}_{A'}) & =
& 2i\zeta(\pi^{A'}\dot{\eta}_{A'} -
\bar{\pi}^A\dot{\bar{\eta}}_A), \nonumber \\
\bar{\pi}{}^A\dot{\bar{\pi}}_A - \pi^{A'}\dot{\pi}_{A'} & = & 0.
\label{IntegConds}
\end{eqnarray}
We find that the integrability conditions (\ref{IntegConds})
coincide with the null string equations of motion in the system
(\ref{RMEqns}).

The conclusion of the subsection is that the integrability
conditions do not contribute additional constraints to the null
string equations of motion~(\ref{RMEqns}). Moreover, the complete
system of the null string equations of motion consists of the
spinor representations for $\dot{x}{}^a$ and $\acute{x}{}^a$ in
(\ref{RMEqns}) together with  their compatibility
conditions~(\ref{ComCon}).

It is easy to show that non-geodesic null string equations of
motion derived in the article~\cite{Ilienko2-Zheltukhin} can be
cast into the form (\ref{RMEqns}). This proves that the two
variational formulations are equivalent on the classical level.
All the results on the properties of those equations also hold in
our case. For more details we refer an interested reader to that
paper. It is also remarkable that the present formulation is free
of the pair of artificial auxiliary world-sheet quantities,
$\rho^\mu$, in the action principle, which was proposed by Bandos
and Zheltukhin in Ref.~\cite{Bandos-Zheltukhin2} and studied by
Zheltukhin and Ilyenko in
Refs.~\cite{Zheltukhin,Ilienko1,Ilienko2-Zheltukhin,Disser}. The
action principle of \cite{Ilienko2-Zheltukhin} contains eight
arbitrary functions of $\tau$ and $\sigma$, namely, two
$\rho^\mu$s and six components of the external antisymmetric gauge
field $B_{ab}(x)$. Nevertheless, as we showed there, only two
gauge invariant combinations of the field strength components of
$B_{ab}(x)$ enter the equations of motion. This means that only
four functions define generic null string dynamics in 4D~Minkowski
space-time, as is the case with the present formulation.

Summarizing, in this section we have shown that the twistor action
functional (\ref{Action}) describes the null string as a
two-dimensional submanifold of 4D~Minkowski space-time with a
degenerate induced metric.
\section{Evolution equation}
\label{Sec.III}%
\subsection{Preliminary results}
\indent%
The invariance of the null string equations of motion under the
restricted gauge transformations of the form (\ref{Scale_Add})
with real functions $q$ and $p$ enables us to impose one more
gauge condition on the functions entering the complete equations
of motion of a null string. It will prove convenient in the
non-geodesic case to fix the extents of the null directions
tangent to the null string world-sheet by imposing the so-called
natural parametrization
\begin{equation}
\dot{\bar{\pi}}_A\bar{\pi}{}^A = 1.%
\label{Nat-Par}
\end{equation}
This amounts to taking $r = \kappa^{-1}$, where $\kappa$ is the
spin-coefficient, whose non-vanishing property shows
that a null congruence is non-geodesic (cf. %
\cite[Vol. 2, p. ??]{SST}). The resulting equations are invariant
under the residual gauge transformations
\begin{equation}
\bar{\pi}{}^A \rightarrow \bar{\pi}{}^A,\,\,\,\, \bar{\eta}{}^A
\rightarrow \bar{\eta}{}^A + p\bar{\pi}{}^A.%
\label{Res-Gauge}
\end{equation}
Here $p$ is a real-valued function of $\tau$ and $\sigma$. The
gauge transformations of the form (\ref{Res-Gauge}) correspond to
the freedom inherent in the definition of the flag planes, which
are associated with the flagpole directions $l^a$ tangent to the
null string world-sheet. The condition (\ref{Nat-Par}) fixes
natural parameter, $\tau$, along the integral curves of the vector
field $l^a$, thus restricting the reparametrization invariance
(\ref{TDiff}) to trivial transformations
\begin{equation}
\tau \rightarrow \tau, \,\,\,\, \sigma \rightarrow \varphi
(\sigma). \label{Res-TDiff}
\end{equation}
Under these transformations the flagpole extent, $\kappa^{-1}$,
remains invariant, whereas $\zeta$ changes to
$\acute{\varphi}{}^{-1}\zeta$.

The condition for the natural parametrization (\ref{Nat-Par})
leads to the identity
\begin{equation}
\ddot{\bar{\pi}}_A\bar{\pi}{}^A = 0.
\end{equation}
It follows that $\ddot{\bar{\pi}}{}^A$ is proportional to
$\bar{\pi}{}^A$. Using the definition of the spin-coefficients
$\kappa$, $\varepsilon$ and $\tau\acute{}$ and noting that
$D\equiv l^{a}\nabla_{a}$, we obtain
\begin{equation}
\ddot{\bar{\pi}}{}^A = [\kappa^{-1}D(\varepsilon\kappa^{-1}) +
\kappa^{-2}(\varepsilon^2 + \kappa\tau\,\acute{}\,)]\bar{\pi}{}^A.
\label{Technical_3.5.2}
\end{equation}
In what follows, we shall denote the expression in the square
brackets in the equation (\ref{Technical_3.5.2}) as $U$.  The
natural parametrization (\ref{Nat-Par}) also gives that
\begin{equation}
\bar{\eta}{}^A = -\dot{\bar{\pi}}{}^A%
\label{EtaChoice}
\end{equation}
up to addition of real multiples of $\bar{\pi}{}^A$.
\subsection{Derivation of equation}
\indent%
We use the spinor representations in (\ref{RMEqns}) for the vector
fields $\dot{x}{}^a$ and $\acute{x}{}^a$ to find a non-linear
evolution equation obeyed by the coordinate, $x^a(\tau,\sigma)$,
of the null string world-sheet. Since the analysis here is
applicable only to the non-geodesic case, we shall restrict our
derivation to the situation when the natural parametrization is
applied.

Let us substitute (\ref{EtaChoice}) and $r=\kappa^{-1}$ into the
definitions of $\dot{x}{}^a$ and $\acute{x}{}^a$ in
(\ref{RMEqns}), the result reads
\begin{equation}
\dot{x}{}^a = \kappa^{-1}\bar{\pi}{}^A\pi^{A'},\,\,\,\,
\acute{x}{}^a = i\zeta(\dot{\bar{\pi}}{}^A\pi^{A'} -
                       \bar{\pi}{}^A\dot{\pi}{}^{A'}).
\label{Technical_3.5.3}
\end{equation}
First, taking the $\tau$-derivative of $\acute{x}{}^a$, we
calculate with the aid of (\ref{Technical_3.5.2}):
\begin{equation}
x\!\!\dot{}\,\acute{}\,{}^a = 
i\dot{\zeta}(\dot{\bar{\pi}}{}^A\pi^{A'} -
\bar{\pi}^A\dot{\pi}^{A'}) + i\zeta(U -
\bar{U})\bar{\pi}{}^A\pi^{A'}.%
\label{XDotPrime}
\end{equation}
Second, taking the $\tau$-derivative of $\dot{x}^a$, we obtain
\begin{equation}
\ddot{x}^a = -\kappa^{-2}\dot{\kappa}\bar{\pi}{}^A\pi^{A'} +
\kappa^{-1}(\dot{\bar{\pi}}{}^A\pi^{A'} +
\bar{\pi}{}^A\dot{\pi}^{A'}), \ddot{x}{}^2 =
-2\kappa^{-2}.%
\label{Support_1}
\end{equation}
One can also find that
\begin{equation}
\acute{x}{}^2 = -2\zeta^2, \,\,\,\, \acute{x}x\!\!\dot{}\,\acute{}
= -2\zeta\dot{\zeta}, \,\,\,\,
\ddot{x}x\!\!\dot{}\,\dot{}\,\acute{} =
2\kappa^{-3}\acute{\kappa}. \label{Support_2}
\end{equation}
Here we have made the use of the identity
$(\ddot{x}{}^2)\acute{}\,=
2\ddot{x}x\!\!\dot{}\,\dot{}\,\acute{}$. In order to obtain the
$\sigma$-derivative of $\kappa$ we can employ the compatibility
conditions of Sec.~\ref{ComConds} with the necessary substitution
$\bar{\eta}{}^A \mapsto -\dot{\bar{\pi}}{}^A$. Taking the
$\sigma$-derivative of $\dot{x}{}^a$ and using
(\ref{Technical_3.5.3}), we derive
\begin{equation}
x\!\!\dot{}\,\acute{}\,{}^A =
-\kappa^{-2}\acute{\kappa}\bar{\pi}{}^A\pi^{A'} +
\kappa^{-1}(\acute{\bar{\pi}}{}^A\pi^{A'} +
\bar{\pi}{}^A\acute{\pi}{}^{A'}).
\end{equation}
Since $(\dot{x}{}^a)\acute{}$ equals to $(\acute{x}{}^a)\dot{}$,
we must also have
$\dot{\bar{\pi}}_A\dot{\pi}_{A'}(\acute{x}{}^{AA'})\dot{}$ $=$
$\dot{\bar{\pi}}_A\dot{\pi}_{A'}(\dot{x}{}^{AA'})\acute{}\,$. This
entails
\begin{equation}
-\kappa^{-2}\acute{\kappa} +
\kappa^{-1}(\dot{\bar{\pi}}_A\acute{\bar{\pi}}{}^A +
\dot{\pi}_{A'}\acute{\pi}{}^{A'}) = i\zeta(U - \bar{U}),
\label{Technical_3.5.4}
\end{equation}
where we used the normalization condition (\ref{Nat-Par}).
Employing the same normalization condition again we find
\begin{equation}
\dot{\bar{\pi}}_A\acute{\bar{\pi}}{}^A - \bar{\pi}_A\bar{\pi}
\!\!\dot{\vphantom{\bar{\pi}}}\hspace{1.6ex}\acute{\vphantom{\bar{\pi}}}\,\,\,{}^A
= 0. \label{Technical_3.5.5}
\end{equation}
On the other hand, taking into account (\ref{Technical_3.5.2}),
the first motion equation in (\ref{RMEqns}) gives
\begin{equation}
\bar{\pi}_A\acute{\bar{\pi}}{}^A + \pi_{A'}\acute{\pi}{}^{A'} = 0.
\end{equation}
Differentiating this equation with respect to $\tau$ and making
the use of (\ref{Technical_3.5.5}) we obtain
\begin{equation}
2(\dot{\bar{\pi}}_A\acute{\bar{\pi}}{}^A +
\dot{\pi}_{A'}\acute{\pi}{}^{A'}) =
(\bar{\pi}_A\acute{\bar{\pi}}{}^A +
\pi_{A'}\acute{\pi}{}^{A'})\dot{}\,=0.
\end{equation}
The substitution of this result into the equation
(\ref{Technical_3.5.4}) finally  yields
\begin{equation}
\acute{\kappa} = -i\kappa^{2}\zeta(U - \bar{U}).
\label{KappaPrime}
\end{equation}

In a generic situation of a non-geodesic case neither $\kappa$ nor
$\zeta$ are equal to zero. Let us then multiply the equation
(\ref{XDotPrime}) by $\kappa^{-2}\zeta^{2}$; the results
(\ref{Technical_3.5.3}), (\ref{Support_1}), (\ref{Support_2}) and
(\ref{KappaPrime}) can be used to show that the evolution equation
has the form
\begin{equation}
\ddot{x}^2\big[\acute{x}{}^2x\!\!\dot{}\,\acute{}\,{}^a -
(\acute{x}x\!\!\dot{}\,\acute{}\,)\acute{x}{}^a\big] -
\acute{x}^2(\ddot{x}x\!\!\dot{}\,\dot{}\,\acute{}\,)\dot{x}{}^a =
0. \label{EvEqn}
\end{equation}
This equation is accompanied by the two constraints
(\ref{OGauge}). A straightforward but tedious calculation shows
that the evolution equation and the constraints are invariant
under the world-sheet reparametrizations of the form
(\ref{TDiff}).

The evolution equation (\ref{EvEqn}), obeyed by the coordinate
$x^a(\tau,\sigma)$ of the non-geodesic null string world-sheet, is
a non-linear counterpart of the free (geodesic) null string
evolution equation
\begin{equation}
\ddot{x}{}^a = 0.%
\label{SGeodEvEqn}
\end{equation}%
\section{Wave-front caustic}
\label{Sec.IV}%
\indent%
Here we present an example of a non-geodesic null two-surface and
explore its connections with the null string interpretation of
such surfaces in 4D~Minkowski space-time and the evolution
equation derived for the non-geodesic null two-surfaces in the
previous section. We shall build this example as a two-dimensional
caustic of a suitable null hypersurface in 4D~Minkowski
space-time.
\subsection{Null hypersurface}
\begin{figure}[b!]
\centering %
\scalebox{1}[1]{\includegraphics[190,147][385,543]{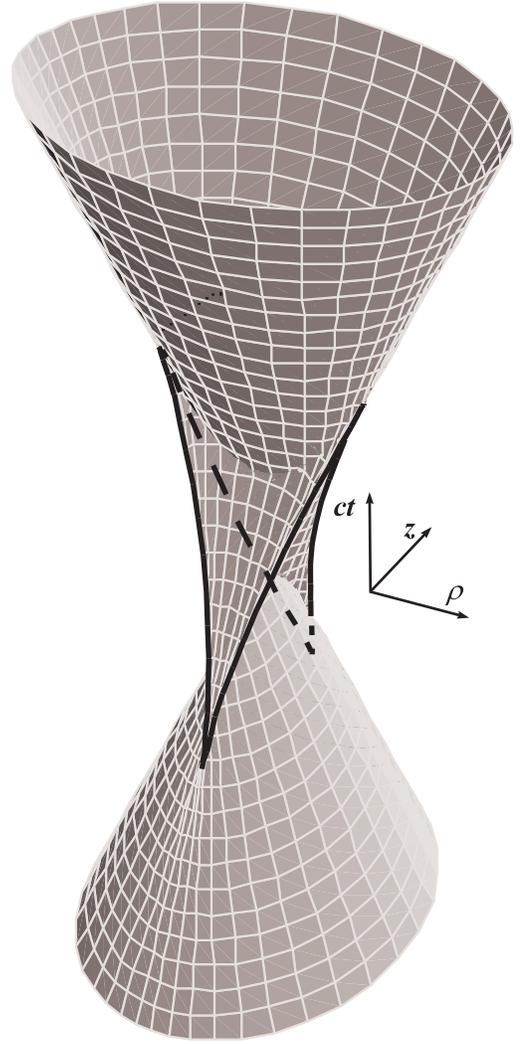}}%
\caption{\label{HSurf} The null hypersurface~(\ref{Hyp_Par}). One
dimension is suppressed. The dark solid and dashed lines show its
two-dimensional null caustic surface.}
\end{figure}

\indent%
Let us start with considering the hypersurface given
parametrically:
\begin{equation}
x^a(u,v,w)%
= \frac{a}{4}\left[
\begin{array}{c}
\hspace{5mm}u \\ (1 + 2\sin^2 v + u)\cos v\cos w \\ (1 + 2\sin^2 v
+ u)\cos v\sin w \\ (3 - 2\sin^2 v - u)\sin v \hspace{10mm}
\end{array}
\right].%
\label{Hyp_Par}
\end{equation}
Here $a$ is a constant with the dimension of length. Taking the
advantage of an axial symmetry present in our example, we
schematically draw the hypersurface using coordinates
($\rho$=$\sqrt{x^2+y^2}$, $z$, $ct$) in Fig.~\ref{HSurf}. Denoting
$x^a_u =
\partial x^a /
\partial u$, $x^a_v =
\partial x^a /
\partial v $ and $x^a_w = \partial x^a / \partial w$, we can
calculate the vector fields spanning the tangent space to this
hypersurface. The result reads
\begin{eqnarray}
x^a_u & = & \frac{a}{4}[1, \cos v \cos w, \cos v \sin w, -\sin v],
\label{TFields} \\%
x^a_v & = & \frac{a}{4}\,(3\cos 2v - u)[0, \sin v \cos w, \sin v
\sin w, \cos v], \nonumber \\%
x^a_w & = & \frac{a}{4}\,(u + 2 - \cos 2v)\cos v%
[0, -\sin w, \cos w, 0]. \nonumber
\end{eqnarray}
For the Lorentz norms of the vector fields we have
\begin{eqnarray}
x^2_u & = & 0,\,\,\,\,x^2_v = -\frac{a^2}{16}\,(u - 3\cos 2v)^2, \nonumber \\%
x^2_w & = & -\frac{a^2}{16}\cos^2 v(u + 2 - \cos 2v)^2.
\end{eqnarray}
One can also check the orthogonal property of these vector fields
in the Lorentz norm $x_u x_v$ = $x_u x_w$ = $x_v x_w$ = $0$. The
equations above show that the hypersurface (\ref{Hyp_Par}) is
null. We can also write the necessarily degenerate induced metric,
$G_{\mu\nu}$. Here the subscript indices $\mu$ and $\nu$ run over
$u$, $v$ and $w$. By definition, it is given by $G_{\mu\nu}$ =
$\partial_\mu x^a \partial_\nu x_a$. Writing $G_{\mu\nu}$ as a
three by three matrix we obtain
\begin{widetext}
\begin{equation}
G_{\mu\nu} = - \frac{a^2}{16}\, \mbox{diag}\,[0, (u - 3\cos 2v)^2,
\cos^2 v (u + 2 - \cos 2v)^2].
\end{equation}
\end{widetext}
This matrix has the rank of two everywhere with
the exception of the following parameter values:
\begin{equation}
u = 3\cos 2v,\,\,\,\,\cos v = 0\,\,\,\,\mbox{or}\,\,\,\,u = \cos
2v - 2. \label{Technical_3.6.1}
\end{equation}
It is of rank one matrix there excluding two points $x^a = (-3a/4,
0, 0,\pm a)$, where the induced metric is of zero rank. One can
show that parameter values $\cos v = 0$ determine two null
straight lines
\begin{equation}
ct = \frac{a}{4} \pm z, \,\,\,\, x = y = 0
\end{equation}
and the parameter values given by the last equality in
(\ref{Technical_3.6.1}) correspond to the segment $|z| \leq a$
contained by the null plain curve
\begin{equation}
ct = -\frac{1}{4a}\,(2z^2 + a^2).
\end{equation}

On the contrary, the first equality in (\ref{Technical_3.6.1})
defines a two-dimensional surface. Substituting $u = 3\cos 2v$
into (\ref{Hyp_Par}), we find for the space-time points belonging
to it:
\begin{equation}
\tilde{x}^a = a\,[\frac{3}{4}\cos 2v, \cos^3 v \cos w, \cos^3 v
\sin w, \sin^3 v].%
\label{NTwo_Par}
\end{equation}
By construction, this is a caustic two-surface for the null
hypersurface (\ref{Hyp_Par}) and the tangent plains to the
two-surface span the null hypersurface. Accounting for the axial
symmetry of our example, the caustic null two-surface is shown by
thick solid and dashed lines in Fig.~\ref{HSurf}.

The new tangent fields at the two-surface are
\begin{eqnarray}
\tilde{x}^a_v & = & -\frac{3}{2}\,a\sin 2v[1, \cos v \cos w, \cos
v \sin w,
-\sin v] \nonumber \\%
\tilde{x}^a_w & = & \hphantom{-\frac{3}{2}}\,a\cos^3 v[0, -\sin w,
\cos w, 0].
\end{eqnarray}
The latter new tangent field coincides with the field obtained as
the result of substitution $u = 3\cos 2v$ in the expression for
$x^a_w$ in (\ref{TFields}), while the former is only a multiple of
the null vector field $x^a_u$ in (\ref{TFields}). For this reason
and to simplify subsequent calculations, we can equally use the
null vector field $x^a_u$ in order to find the spinor
corresponding to the null direction tangent to the two-surface.
Since the first tangent vector field $\tilde{x}^a_v$ has zero
Lorentz norm and the second tangent vector field $\tilde{x}^a_w$
is space-like at the points given by the equation
(\ref{NTwo_Par}), we infer that the two-surface is a null
two-surface in 4D Minkowski space-time. The parametric
representation (\ref{NTwo_Par}) also provides a representation for
this surface as an intersection of two hypersurfaces:
\begin{eqnarray}
x^2 + y^2 + z^2 & = & \frac{a^2}{4} \left[1 +
3\left(\frac{4ct}{3a}\right)^2\right], \label{Technical_3.6.2} \\%
x^2 + y^2 - z^2 & = & \frac{a^2}{4} \left(\frac{4ct}{3a}\right)
\left[3 +
\left(\frac{4ct}{3a}\right)^2\right]. \nonumber%
\end{eqnarray}
Here $ \left|4ct/3a\right|$ $\leq$ $1$. Eliminating $4ct/3a$ from
the equations in (\ref{Technical_3.6.2}), we observe that in a
particular reference frame the projection of this null two-surface
in a hyperplane of constant time is an astroid of revolution with
the parameter $a$ given by the equation:
\begin{equation}
[x^2 + y^2 + z^2 - a^2]^3 + 27a^2[x^2 + y^2]z^2 = 0
\end{equation}
(see Fig.~\ref{Strings}).

As well known, a compact space-like two-surface in 4D~Minkowski
space-time can be represented as an intersection of two null
hypersurfaces. Somewhat analogues to that situation,
two-dimensional self-intersections (caustics) of null
hypersurfaces provide examples  of non-geodesic null two-surfaces
in 4D~Minkowski space-time.

In order to make connection with the description of the previous
sections, we need the spin-tensor expressions of various vector
field quantities.
As mentioned above, the null vector field $x^a_u$ can be employed
to obtain the spinor field describing the null directions tangent
to the null two-surface. First, we explicitly calculate
\begin{equation}
x^{AA'}_u [\,= x^a_u (\sigma_a)^{AA'}] = \frac{a}{4\sqrt{2}}
\left[
\begin{array}{ll}
1 - \sin v              & \cos v \,\mbox{e}^{iw} \\ \cos v
\,\mbox{e}^{-iw} & 1 + \sin v
\end{array}
\right].%
\label{Technical_3.6.3}
\end{equation}
[Here $(\sigma_a)^{AA'}$ are the Pauli matrices.] Since the vector
field $x^a_u$ is real-valued and null, the determinant of the
matrix in (\ref{Technical_3.6.3}) vanishes and we must have
$x^{AA'}_u \propto \bar{\pi}{}^A\pi^{A'}$. The components of the
spinor $\bar{\pi}_A$ can be taken as follows:
\begin{equation}
\bar{\pi}_A = -i\sqrt{2} \left[
\begin{array}{r}
\cos\left(\frac{v}{2} - \frac{\pi}{4}\right)\mbox{e}^{-iw/2} \\
\sin\left(\frac{v}{2} -
\frac{\pi}{4}\right)\mbox{e}^{\,\hspace{2mm}iw/2}
\end{array}
\right].%
\label{Technical_3.6.5}
\end{equation}
\subsection{Non-geodesic null string}
\indent%
Now we are in a position to explore connections between the
example null two-surface and a non-geodesic null string
world-sheet in 4D~Minkowski space-time,
\cite{Ilienko2-Zheltukhin}. Changing the parameters $(v,w)$ to
$(\tau,\sigma)$ we can rewrite the parametric representation
(\ref{NTwo_Par}) for the null two-surface as $x^a$ =
$a$$[(3/4)\cos 2\tau,$ $\cos^3\tau \cos\sigma,$ $\cos^3\tau
\sin\sigma,$ $\sin^3\tau]$. The range of the parameters is $\tau
\in$ $[\pi /2,\pi]$ and $\sigma \in$ $[0,2\pi]$. This corresponds
to a closed null string with the parameter $\tau$ playing the role
of a time variable. The results (\ref{Technical_3.6.3}) and
(\ref{Technical_3.6.5}) allow us to write
\begin{equation}
\dot{x}^{AA'} = \frac{3a}{2\sqrt{2}}\sin
2\tau\,\bar{\pi}{}^A\pi^{A'}.%
\label{TNull}
\end{equation}
Here $\bar{\pi}_A$ is given by the formula (\ref{Technical_3.6.5})
with the necessary change of $(v,w)$ to $(\tau,\sigma)$. Next, we
introduce a second spinor field $\bar{\eta}{}^A$:
\begin{equation}
\bar{\eta}{}^A = \frac{i}{\sqrt{2}} \left[
\begin{array}{l}
\cos\left(\frac{\tau}{2} -
\frac{\pi}{4}\right)\mbox{e}^{\hspace{2mm}i\sigma/2} \\
\,\sin\left(\frac{\tau}{2} -
\frac{\pi}{4}\right)\mbox{e}^{-i\sigma/2}
\end{array}
\right].%
\label{ASBasis}
\end{equation}
which, together with $\bar{\pi}_A$, constitutes a normalized
New\-man-Pen\-ro\-se dyad (spin-frame) for all admissible values
of the parameters $\tau$ and $\sigma$. Making the use of
(\ref{Technical_3.6.5}) and (\ref{ASBasis}) we obtain
\begin{equation}
\acute{x}{}^{AA'} = -i\frac{a}{\sqrt{2}}\,\cos^3\tau\,
                    (\bar{\pi}{}^A\eta^{A'} - \bar{\eta}{}^A\pi^{A'}).
\label{TSpace}
\end{equation}

The vector fields $\dot{x}{}^a$ and $\acute{x}{}^a$ tangent to the
null two-surface vanish at two space-time points
\begin{equation}
ct = -\frac{3a}{4},\,\,\,\, x = y = 0,\,\,\,\, z = \pm\,a
\label{BadPoints}
\end{equation}
and, in addition, the vector field $\dot{x}{}^a$ vanishes on the
circle
\begin{equation}
ct = \frac{3a}{4},\,\,\,\, x^2 + y^2 = a^2,\,\,\,\, z = 0.
\label{BadCircle}
\end{equation}

Comparing the results (\ref{TNull}) and (\ref{TSpace}) with the
first two equations in the system (\ref{RMEqns}), we can identify
the functions $r$ and $\zeta$ as follows:
\begin{equation}
r = \frac{3a}{2\sqrt{2}}\,\sin 2\tau, \,\,\,\, \zeta =
-\frac{a}{\sqrt{2}}\,\cos^3\tau.%
\label{Herr}
\end{equation}
The expressions for $\bar{\pi}{}^A$ and $\bar{\eta}{}^A$ can be
used to obtain the functions $(\omega - \ln|\zeta|)\dot{}$ and
$\psi$ of the Ref.~\cite{Ilienko2-Zheltukhin} directly, the result
reads:
\begin{equation}
(\omega - \ln|\zeta|)\dot{} = 0,\,\,\,\, \psi =
\frac{a}{4\sqrt{2}}\,.%
\label{Technical_3.6.6}
\end{equation}
This knowledge is important for relating $\mbox{Re}\,\dot{\omega}$
and $\psi$ with the quantities which represent the field strength
of the gauge field $B_{ab}(x)$ of the
paper~\cite{Ilienko2-Zheltukhin}. Thus, we obtain
\begin{eqnarray}
\psi                    & = &
-\frac{3}{4}\,\mbox{\ae}a^2\sin2\tau\cos^3\tau(\rho^\tau)^{-1}\phi,
\nonumber \\%
2\mbox{Re}\,\dot{\omega} & = &
-\frac{3}{2}\,\mbox{\ae}a^2\sin2\tau\cos^3\tau(\rho^\tau)^{-1}
\mbox{Re}\,\nu \\%
& & + \left(\ln|\cos^6\tau\,(\rho^\tau)^{-1}|\right)^.. \nonumber
\end{eqnarray}
Substituting the result (\ref{Technical_3.6.6}) in these equations
we have
\begin{eqnarray}
\phi & = & -\,\rho^\tau
[3\sqrt{2}\mbox{\ae}a\sin2\tau\cos^3\tau]^{-1}, \\
\mbox{Re}\,\nu & = &
-\,\rho^\tau[\frac{3}{2}\,\mbox{\ae}a^2\sin2\tau\cos^3\tau]^{-1}
\left(\ln|\cos^6\tau\,(\rho^\tau)^{-1}|\right)^.. \nonumber
\end{eqnarray}
The formulae above show that the quantity $\phi$ representing the
only physical degree of freedom of the field strength diverges at
the space-time points (\ref{BadPoints}) and on the circle
(\ref{BadCircle}).

\begin{figure}[t!]
\centering %
\scalebox{1}[1]{\includegraphics[168,236][406,718]{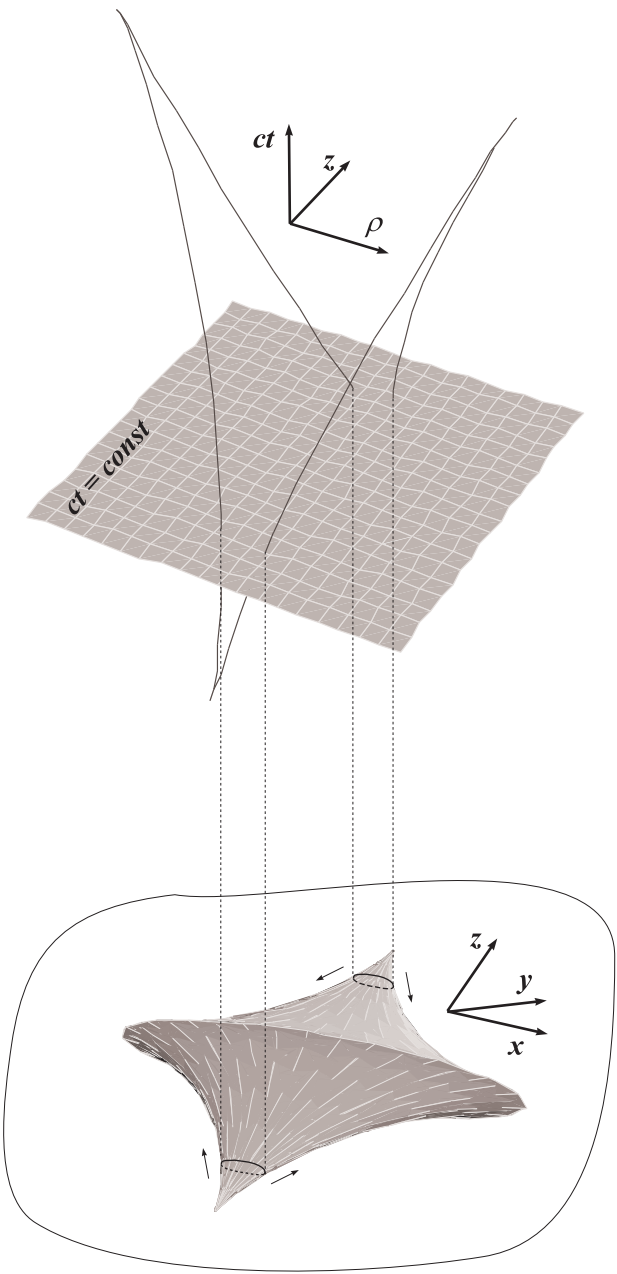}}%
\caption{\label{Strings} Null string interpretation of the
two-dimensional caustic surface. (a) A section of the two-surface
by a hyperplane of constant time (one space dimension is not
shown). (b) The space picture of the null strings dynamics.}
\end{figure}

Therefore, in a particular reference frame, Fig.~\ref{Strings},
one can interpret the null two-surface of this section as a pair
of circular null strings which appear with zero radius at $t =
-3a/4c$ at spatial points ($x$, $y$, $z$) = ($0$, $0$, $\pm a$).
They then expand until the time $t = 3a/4c$, when they disappear
having the circumference of $2\pi a$. The null strings are the
sections of the caustic two-surface in Fig.~\ref{HSurf} by the
hyperplanes of constant time. Projection of the world-sheets of
the null strings into a particular reference frame constitutes the
astroid of revolution described earlier in this section.
\subsection{Evolution equation}
\indent%
It is also interesting to make connections of this example with
the evolution equation for non-geodesic null strings derived in
the previous section. The parametric expressions for
$\bar{\pi}{}^A$ and $\bar{\eta}{}^A$, together with the results
(\ref{TNull}) and (\ref{TSpace}), can be utilized to verify the
equality (\ref{EtaChoice}).
Noting that $\kappa$ = $r^{-1}$, we find
\begin{equation}
\acute{\kappa} = 0,\,\,\,\, \dot{\zeta} =
\frac{3a}{2\sqrt{2}}\sin2\tau\cos\tau \,\,\,\,\mbox{and}\,\,\,\, U
- \bar{U} = 0.
\end{equation}
Then, we have $x\!\!\dot{}\,\acute{}\,{}^a$ =
$\zeta^{-1}\dot{\zeta}\acute{x}{}^a$. Next, the formulae
(\ref{Support_1}) and (\ref{Support_2}) yield
\begin{eqnarray}
\acute{x}{}^2 & = & -\,a^2\cos^6\tau, \hspace{2.7em}
\ddot{x}x\!\!\dot{}\,\dot{}\,\acute{} =
0, \nonumber \\%
\acute{x}x\!\!\dot{}\,\acute{} & = &
\frac{3}{2}\,a^2\sin2\tau\cos^4\tau,\,\, \ddot{x}{}^2 =
-\frac{9}{4}\,a^2\sin^2 2\tau.
\end{eqnarray}
Finally, with the aid of results obtained above, we observe that
the evolution equation (\ref{EvEqn}) holds.
\section{Discussion and Outlook}
\label{Sec.V}%
\indent%
Firstly, the method employed in this paper for obtaining a
variational principle, (\ref{Action}), for a null two-surface can
be, in principle, used for designing a twistor variational
principle for time-like two-surfaces (conventional strings) and
space-like two-surfaces. The idea is to take a simple bivector
field $p_{ab}(x)$ and impose one of the algebraic conditions
$p_{ab}p{}^{ab}$ = $1$ or $p_{ab}p{}^{ab}$ = $-1$. The latter
condition would single out the string (compare with the last part
of \cite{Gusev-Zheltukhin}), while the former would correspond to
a space-like two-surface. It is easy to see that such a procedure
uniquely fixes the symmetric second rank spin-tensor field
$\phi{}_{AB}(x)$ in the standard decomposition of an antisymmetric
4D~Minkowski space-time tensor field
$p{}_{ab}(x)=\phi{}_{AB}(x)\varepsilon{}_{A'B'} +
\bar{\phi}{}_{A'B'}(x)\varepsilon{}_{AB}$. Then, the variational
principle
\begin{equation*}
S = \frac{1}{2!}\int [\phi{}_{AB}(x)\varepsilon{}_{A'B'} +
\bar{\phi}{}_{A'B'}(x)\varepsilon{}_{AB}]dx^{AA'} \wedge dx^{BB'}
\end{equation*}
would define a two-surface subject to the differential constraint
stated in Sec.~\ref{Intro} [see the equation~(\ref{DiffCon})].
Now, one hopes that the use of spinor decomposition for
$p_{ab}(x)$, consistent with either of the formulated algebraic
constraints, would provide equations of motion, which
automatically incorporate the differential constraints formulated
in the Sec.~\ref{Intro}. Such an assertion is supported by the
success of this procedure for the null two-surfaces (null strings)
presented in the current contribution. It may well be possible to
derive the analogues of the evolution equation (\ref{EvEqn}) for
generic (interacting) strings in 4D~Minkowski space-time and
curved space-times of general relativity, where exist explicit
spinor and twistor constructions (cf. \cite[Eqn. (16)]{Indusy} and
\cite[Eqns. (3.6)--(3.11)]{Flaherty}). In the same way it should
be possible to build twistor action functionals in the both cases
for generic time-like and space-like
two-surfaces of 4D~Minkowski space-time.%

Secondly,\hfill if\hfill one\hfill employs\hfill Feber's\hfill definition\hfill of\hfill a\\
SU$(2, 2\,|\,N)$ supertwistor \cite{Feber},  one could build a
description of null strings with spin in the physical dimensions
of space-time. Its analisys presumably would follow the standard
pass outlined in the works of Shirafuji~\cite{Shirafuji} and
Bengtsson et al.~\cite{Bengtsson_et_al}.
\section*{ACKNOWLEDGEMENTS}
\indent%
I acknowledge an early conversation with A.A.~Zheltukhin on the
matters of the article \cite{Gusev-Zheltukhin}. I am very grateful
to R.~Penrose and Yu.P.~Stepanovskii for their interest in this
work. Special thanks go to T.Yu.~Yatsenko for the help with the
production of pictures.

\end{document}